\documentstyle[aps,prc,preprint]{revtex}

\def\bra#1{\bigl\langle #1\bigr|}
\def\ket#1{\bigl| #1\bigr\rangle}

\begin{document}
\draft
\title{Charged current weak electroproduction of $\Delta$ resonance} 
\author{L. Alvarez-Ruso, S.K. Singh\cite{singhper} 
and M.J. Vicente Vacas}
\address{Departamento de F\'{\i}sica Te\'{o}rica and IFIC, Centro Mixto 
Universidad de Valencia-CSIC,\\ 46100 Burjassot, Valencia, Spain}
\date{\today}
\maketitle

\begin{abstract}
We study the weak production of $\Delta$ (i.e.
$e^{-}+p \rightarrow \Delta ^{0}+ \nu_{e}$ and
$e^{+}+p \rightarrow \Delta ^{++}+ \bar{\nu}_{e}$) in the
intermediate energy range corresponding to
the Mainz and TJNAF electron accelerators. The differential cross
sections $\sigma(\theta)$ are found to be of the order of
$ 10^{-39}$ cm$^2$/sr, over a range of angles which increases
with energy. The possibility of observing  these reactions with the
high luminosities available at these
accelerators, and studying the weak N-$\Delta$ transition form factors
through these reactions is discussed.
The production cross section of N$^*(1440)$
in the kinematic region of $\Delta$ production is also estimated
and found to be small.  
\end{abstract}


\section{Introduction}
The study of nucleon  and its excitation spectrum has been pursued for a 
long time both experimentally and theoretically, specially in the region
of the $\Delta$ resonance. In this region, the QCD inspired quark models for 
baryon structure provide theoretical insight in the nonperturbative regime 
of QCD at low and intermediate energies. At these energies, the study of 
various electromagnetic excitation processes, induced by electrons and
photons,
has been made at many research facilities around the world, and there exists 
extensive literature on electromagnetic transition form factors  
\cite{1}. On the other hand, similar studies on the 
corresponding weak form factors have been few and far between. 
In a series of experiments done with intermediate energy neutrinos 
at ANL, BNL and CERN laboratories, attempts have been made to study the 
transition form factors for the charge changing weak current, and there 
exists a fair amount of data to determine these form factors 
\cite{2,3,4,5,6}.
This is not the case with the neutral current processes, where
there are very few experiments \cite{7,8} in the intermediate 
energy range and no serious analysis has been made of the
transition form factors. The main interest in the neutral current
sector  has been to study the parity violating asymmetry in the 
polarised  electron scattering with nucleons and nuclei in order to explore 
the nonzero strangeness content of the nucleon\cite{9,10}.

Now, the availability of continuous wave electron accelerators with
100\% duty cycle in the energy range of few GeV, and the possibility of
achieving very high luminosities at these accelerators has led to the
feasibility of performing electron scattering experiments in the resonance 
region with very good statistics\cite{11,12}. These experimental
studies, in principle, can be extended to explore weak interaction
physics in the $\Delta$ resonance region. In this paper, we explore the 
possibility of doing such experiments and present a quantitative analysis 
of the charged current reaction in which $\Delta$'s are produced. Similar 
theoretical studies in the neutral current sector have been performed
earlier\cite{13} and most recently by Mukhopadhyay {\it et al}\cite{14}.

The aim of the present paper is to give a general analysis of the weak 
production of $\Delta$ through the processes 

\begin{mathletters}
\begin{equation}
e^{-}+p \rightarrow \Delta ^{0}+ \nu_{e}
\end{equation}
and
\begin{equation}
e^{+}+p \rightarrow \Delta ^{++}+ \bar{\nu}_{e}\,,
\end{equation}
\end{mathletters}

\noindent
and to examine critically the feasibility of doing such an experiment at the
Mainz Microtron and/or at TJNAF. These reactions were earlier studied by
Hwang {\it et al.} \cite{15}, where a bag model was used to calculate the
N-$\Delta$ form factors and the effect of the $\Delta$ width was not taken 
into account.
In this work we retain all the weak vector and axial vector form factors in
the matrix elements of N-$\Delta$ transitions.
The present available information on these form factors from the experimental 
data on electromagnetic and neutrino
production of $\Delta$ has been fully utilized through the application of
Conserved Vector Current (CVC)and Partially Conserved Axial Current (PCAC)
hypotheses in the N-$\Delta$ sector
\cite{16,17,18,19,20}. In addition, the width of $\Delta$ resonance is properly
taken into account and is found to give important effects on the differential
cross section $\sigma(\theta)$. The effect of various parametrizations of the
N-$\Delta$ form factors, discussed recently in the literature \cite{16,17}, 
has been studied to explore the possibility of distinguishing between them 
experimentally. Finally, we have also estimated the production cross section 
for the Roper, N$^*$(1440), the next higher resonance, in order to 
understand its effect in the kinematic region
of $\Delta$ production. We find its effect to be sufficiently small and well 
separated from the kinematic region of present interest to allow for a clean 
identification of $\Delta$ through observation of the pions and nucleons
produced as decay products.

In section II, we describe the transition currents for the production of
$\Delta$ and N$^*$(1440), and derive expressions for the cross sections. 
In section III, we present the numerical results for the differential cross 
sections $\sigma(\theta)$ for the considered reactions and discuss the 
possibility of experimentally observing them, in section IV.

\section{Transition Currents and Cross sections}

\subsection{$e^{-}+p \rightarrow \Delta ^{0}+ \nu_{e}$ and
$e^{+}+p \rightarrow \Delta ^{++}+ \bar{\nu}_{e}$}

The matrix element for the process 
$e^{-}(k)+p(p) \rightarrow \Delta ^{0}(p')+ \nu_{e}(k')$
is written as \cite{21}

\begin{equation}
{\cal M}= {{G}\over{\sqrt{2}}}\, \cos \theta_{c}\, l_{\alpha} J^{\alpha} \,,
\end{equation}
with

\begin{equation}
 l_{\alpha}= \bar{u}(k') \gamma_{\alpha} (1 -\gamma_{5}) u(k)\,,
\end{equation}
and
\begin{eqnarray}
 \nonumber 
 J^{\alpha}=& \bar{\psi_{\mu}}(p') \{ [
  {{C_3^V}\over{M}} (g^{\mu \alpha} q \!\!\! / - q^{\mu} \gamma^{\alpha})+
  {{C_4^V}\over{M^2}} (g^{\mu \alpha} q\cdot p' - q^{\mu} p'^{\alpha}) 
  \\ \nonumber 
  &+ {{C_5^V}\over{M^2}} (g^{\mu \alpha} q\cdot p - q^{\mu} p^{\alpha})
  ] \gamma_{5}\\ 
  &+ {{C_3^A}\over{M}} (g^{\mu \alpha}q \!\!\! / - q^{\mu} \gamma^{\alpha})+
 {{C_4^A}\over{M^2}} (g^{\mu \alpha} q\cdot p' - q^{\mu} p'^{\alpha})+
 {C_5^A} g^{\mu \alpha}\\ 
 \nonumber 
  &+ {{C_6^A}\over{M^2}} q^{\mu} q^{\alpha} \} u(p)\,,
\end{eqnarray}  
where $M$ is the nucleon mass, $\psi_{\mu}(p')$ and $u(p)$ are the 
Rarita Schwinger and Dirac spinors for $\Delta$ and nucleon of momentum $p'$ 
and $p$, $q=p'-p=k-k'$, is the momentum transfer, $C_i^V$ and $C_i^A$ 
($i=3,4,5,6$) are the vector and axial vector transition form factors as 
defined by Llewellyn Smith \cite{20} and are discussed in detail in 
section II.B.
It is relevant to mention here that $C_6^V(q^2) \equiv 0$, assuming CVC. 
With the matrix element given in Eqs. (2)-(4),
the differential cross section $d \sigma/ d \Omega_{\Delta}$ is calculated 
to be

\begin {equation}
{{d \sigma}\over{ d \Omega_{\Delta}}} = 
 \frac{1}{16 \pi^3}\, G^2 \cos^2\theta_c \int d|\bbox{p}'|
\frac{|\bbox{p}'|^2}{E_e E_\nu} \frac{\Gamma/2}{(W - M')^2 + 
\Gamma^2/4} L_{\alpha \beta} J^{\alpha \beta}\,,
\end {equation}
with

\begin {equation}
L_{\alpha \beta} = k_\alpha k'_\beta + k_\alpha' k_\beta
- g_{\alpha \beta} k . k' + i \epsilon_{\alpha \beta
\gamma \delta} k^\gamma k'^{\delta}\,,
\end {equation}
and

\begin {equation}
J_{\alpha \beta} = \bar{\Sigma} \Sigma J_\alpha^\dagger J_\beta\,,
\end {equation}
where the summation is performed over the hadronic spins using a spin $3/2$
projection operator $P_{\mu \nu}$ given by

\begin {equation}
P_{\mu \nu} = -\frac{p' \!\!\!\! / + M'}{2 M'} \left( g_{\mu
\nu}-\frac{2}{3}\frac{p'_{\mu}p'_{\nu}}{M'^2} +
\frac{1}{3}\frac{p'_{\mu}\gamma_{\nu} - p'_{\nu}\gamma_{\mu}}{M'}
-\frac{1}{3}\gamma_{\mu}\gamma_{\nu} \right)
\end{equation} 
In Eq. (5), $W$ is the invariant mass of the $\Delta$ given by
$W^2=p'^2$, $M'$ is the  $\Delta$ mass and $\Gamma$ is its decay width
given by

\begin{equation}
\Gamma= {\frac{1}{6\pi}} \left ( {\frac {f^*}{m_{\pi}}} \right )^2
        {\frac{M}{W}}\, |\bbox{q}_{cm}|^3 \theta(W-M-m_{\pi})\,,
\end{equation}
where $m_{\pi}$ is the pion mass, ${\bbox{q}_{cm}}$ is the pion momentum in
the rest frame of the resonance  and $f^*=2.13$. 

We now turn to the process $e^{+}+p \rightarrow \Delta ^{++}+ \bar{\nu}_{e}$.
The matrix element is written in the same way as Eqs. (2)-(4) with
the following replacements:

(i) The leptonic current $l_\alpha$ in Eq. (3), now involves antiparticles
and is written in terms of $v$ spinors instead of $u$ spinors,
 
(ii) The matrix element of the hadronic current $J_{\alpha}$ in Eq. (4)
is now evaluated between initial proton and final $\Delta ^{++}$
states, with the relation

\begin {equation}
{\bra{\Delta ^{++}}} J_{\alpha} {\ket{p}} =\sqrt{3}\,
{\bra{\Delta^{0}}} J_{\alpha} {\ket{p}}.
\end {equation}
With these two changes the differential cross section is effectively given 
by Eq. (5), with $L_{\alpha\beta}(k,k') \rightarrow L_{\alpha\beta}(k',k)$
and  $J_{\alpha\beta}(p,p') \rightarrow 3J_{\alpha\beta}(p,p')$

\subsection{N-$\Delta$ transition form factors}

The N-$\Delta$ transition form factors relevant to the weak transition current 
have been discussed in the literature in connection with the analysis 
of neutrino scattering experiments\cite{18,19,20,21} and in the 
context of quark model calculations\cite{16,17}. We summarise in this section 
some details of these form factors, needed for present calculations.

\subsubsection {Vector form factors}

As stated in section II.A, there are four weak vector form factors
$C_3^V$, $C_4^V$, $C_5^V$  and $C_6^V$ occurring in this transition. The
imposition of the CVC hypothesis implies $C_6^V=0$. The other three form
factors are then given in terms of the isovector electromagnetic form
factors in the p-$\Delta^+$ electromagnetic transition. 
Specifically, the hadronic matrix element for the reactions (1a) and
(1b) are given as

\begin{mathletters}
\begin{equation}
\bra {\Delta^{0}} J_{\alpha} \ket{p} = \bra {\Delta^{+}} J_{\alpha}^{em}(T=1) 
\ket{p}
\end{equation}
and 
\begin{equation}
\bra {\Delta^{++}} J_{\alpha} \ket{p} = \sqrt{3}\, \bra{\Delta^{+}} 
J_{\alpha}^{em}(T=1) \ket{p}\,,
\end{equation}
\end{mathletters}
where $J_{\alpha}^{em}(T=1) $ is the isovector electromagnetic current.

The information on the isovector electromagnetic form factors 
$C_i^V(q^2)\; (i= 3,4,5)$ is obtained from the analysis of photo and 
electroproduction data of $\Delta$, which is done in terms of the multipole 
amplitudes $E_{1+},M_{1+}$ and $S_{1+}$ \cite{22}. The present data on 
$E_{1+}$ and $S_{1+}$ amplitudes are very meager and these amplitudes
are expected to be small. Assuming $M_{1+}$ dominance of the 
electroproduction amplitude, which is predicted by the nonrelativistic 
quark model, the form factors $C_i^V(q^2)$ satisfy the relations:

\begin{equation}
C_5^V=0,\;\;\; C_4^V=-{{M}\over{M'}}C_3^V\,.
\end{equation}
The relations given in Eq. (12) have been used in the analysis of the 
electroproduction experiments and $C_3^V(q^2)$ has been determined. The 
following parametrizations of $C_3^V(q^2)$, available in the literature
\cite{16,17} are used in our present calculations:

\begin{eqnarray}
&&1.\;\;\;C_3^V(q^2) = {{2.05}\over{(1-q^2/0.54{\rm GeV}^2)^2}}\,, \\
&& \nonumber \\
&&2.\;\;\;C_3^V(q^2) = {{1.39}\over{
\sqrt{1-q^2/1.43{\rm GeV}^2)}(1-q^2/0.71{\rm GeV}^2)^2}} 
\end{eqnarray}
For the purpose of comparison with a simple form factor obtained in the 
quark model, we also use \cite{16}

\begin{eqnarray}
&&3.\;\;\;C_3^V(q^2) = {{M}\over{\sqrt{3}m}} e^{-{\bar q}^2/6}
\end{eqnarray}
where $m = 330$ MeV is the quark mass and $\bar q= |\bbox{q}|/ \alpha_{HO}$,
with $\alpha_{HO}=320$ MeV, being the harmonic oscillator parameter. 

\subsubsection {Axial Vector form factors}

 There are four axial vector  form factors $C_3^A$, $C_4^A$, $C_5^A$ 
and $C_6^A$, as defined in Eq. (4). Using the Pion pole Dominance
of the Divergence of Axial Current (PDDAC), $C_6^A(q^2)$ can be given 
by the equation \cite{19}

\begin{equation}
C_6^A(q^2) = {{g_{\Delta} f_{\pi}}\over{2 \sqrt{3} M}}
 {{M^2}\over{m_{\pi}^2-q^2}}\,,
\end{equation}
where $g_{\Delta}=f^*2M/m_\pi$ is the $\Delta^{++}\rightarrow p\,\pi^+$ 
coupling constant and $f_{\pi} = 0.97 m_{\pi}$ is the pion decay constant.
Evaluating the matrix element of the divergence of the axial current
in the limit $m_{\pi}^2 \rightarrow 0$ and $q^2\rightarrow 0$, gives the
off diagonal Goldberger-Treiman relation

\begin{equation}
C_5^A(0) = {{g_{\Delta} f_{\pi}}\over{2 \sqrt{3} M}},\;\;\;
C_6^A(q^2) = C_5^A {{M^2}\over{m_{\pi}^2-q^2}}.
\end{equation}
In absence of any other theoretical input, $C_3^A(q^2)$, $C_4^A(q^2)$ and
$C_5^A(q^2)/C_5^A(0)$ remain undetermined. The data on neutrino scattering
are analysed using these form factors as free parameters and using 
equations (12)-(14) for the vector form factors. The parametrizations used for
the various axial form factors are given below, where dipole form factors
have been modified for a better fit to the data \cite{2,3,4,5,6}

\begin{equation}
C_{i=3,4,5}^A(q^2) = {{C_i(0)\left[ 1-{{a_i q^2}\over{b_i-q^2}} \right] }
{\left( 1- {{q^2}\over{M_A^2}}\right)^{-2}}}
\end{equation}
with $C_3^A(0)=0$, $C_4^A(0)=-0.3$, $C_5^A(0)=1.2$, $a_4=a_5=-1.21$, 
$b_4=b_5=2$ GeV$^2$ and $M_A=1.0$ GeV. 
Recently, these form factors have been calculated in some quark models and 
a comparative study of various models has been 
presented \cite{16}. For a comparison, with the phenomenological form 
factors, we also use a nonrelativistic quark model calculation\cite{16}

\begin{equation}
C_5^A(q^2)=\left(\frac{2}{\sqrt{3}}+\frac{1}{3\sqrt{3}}\frac{q_0}{m}\right) 
e^{-{\bar q}^2/6} \ , \ C_4^A(q^2)=-\frac{1}{3\sqrt{3}}\frac{M^2}{M'm}
e^{-{\bar q}^2/6} \ , \ C_3^A(q^2)=0.
\end{equation}

\subsection{$e^{-}+p \rightarrow N^*+ \nu_{e}$}

The Roper resonance N$^*$, with mass $1440$ MeV and decay width of
about $350$ MeV \cite{23}, is the next higher resonance which has
appreciable strength into $N \pi$ decay channel. The $N \pi$ events coming
from the N$^*$ decay can lie in the invariant mass region of the $\Delta$
resonance. Therefore, we also calculate the production cross section of N$^*$
resonance and the possibility to separate it from a $\Delta$ resonance signal.
It is to be noted that there is no corresponding reaction with an $e^+$
beam thus, the $\Delta^{++}$ production signal is cleaner than the 
$\Delta^{0}$ production signal.

The matrix element for the process $e^-(k) + p(p) \rightarrow N^*(p') +
\nu_{e}(k')$ is written assuming standard properties of the charged weak current
$J^\alpha$ in the $\Delta S = 0$ sector, neglecting second class currents
\cite{24}. Using constraints free form factors and manifestly gauge
invariant operators for the vector current matrix element \cite{25},
$J^\alpha$ is written as 

\begin{equation}
J^\alpha = \bar u_{N^*}(p') \left[ F_1^{V}(q^2)\left( q \!\!\!\!\, /\, 
q^{\alpha} - 
q^2 \gamma^{\alpha}\right) + i F_2^{V}(q^2) \sigma^{\alpha \beta} q_{\beta} +
F_A^{V}(q^2) \gamma^{\alpha} \gamma_5 + F_P^{V}(q^2) q^{\alpha} \gamma_5  
\right] u(p)\,,
\end{equation}

\noindent
where $F_{1,2}^{V}(q^2)$ and $F_{A,P}^{V}(q^2)$ are the isovector
vector and axial vector form factors. 

The expression for the differential cross
section $d\sigma/d\Omega_{N^*}$ is given by Eq. (5) with $M'$ and $\Gamma$
replaced by the N$^*$ mass and its width respectively. For the N$^*$ width we
have chosen the model described in the Appendix of Ref.\cite{26}, where both 
$\pi + N$ and $\pi + \pi + N$ partial decay channels are taken into account. 
The $\pi + \pi + N$ decay is assumed to go through a $\pi + \Delta$ 
intermediate state.  

\subsection{N-N$^*$ transition form factors}

Using the matrix element in Eq. (20) $F_1^{V}(q^2)$ and $F_2^{V}(q^2)$ can, in
principle, be determined from the available experimental data on photo and
electroproduction of Roper resonance from protons and neutrons. The
data on the photoproduction of protons and neutrons fix
only $F_2^{V}(0)$. The electroproduction of Roper resonance has been measured 
only for the proton, and data are not of very good
quality \cite{27}. In absence of any data on the neutron target, we have to
rely on a model to determine the isovector form factors $F_{1}^{V}(q^2)$ and
$F_{2}^{V}(q^2) / F_{2}^{V}(0)$. There are many models in the literature for
the electroproduction of Roper resonance with very different results
\cite{28}. For the purpose of present estimates, we use a simple
nonrelativistic quark model, in which the isovector transverse and
longitudinal helicity amplitudes $A_{\frac{1}{2}}^V(q^2)$ and
$S_{\frac{1}{2}}^V(q^2)$, respectively, satisfy \cite{28}

\begin{equation}
A_{\frac{1}{2}}^{V}(q^2) = \frac{5}{3} A_{\frac{1}{2}}^{p}(q^2) \ $and$ \
S_{\frac{1}{2}}^{V}(q^2) = S_{\frac{1}{2}}^{p}(q^2)
\end{equation}

\noindent
The helicity amplitudes $A_{\frac{1}{2}}^p$ and $S_{\frac{1}{2}}^p$,
the superscript $p$ referring to the proton are defined in the standard 
way \cite{29} i.e.

\begin{eqnarray}
A_{\frac{1}{2}}^p & = & \sqrt{\frac{2 \pi \alpha}{k_R}} {\bra{N^*
\uparrow}} \sum_{pol.} \epsilon \cdot J_{em}^{p} {\ket{N \downarrow}}\,,
\nonumber\\
 & & \\
S_{\frac{1}{2}}^p & = & \sqrt{\frac{2 \pi \alpha}{k_R}}
\frac{|{\bbox q|}}{\sqrt{-q^2}}
{\bra{N^*\uparrow}} \sum_{pol.} \epsilon \cdot J_{em}^{p} {\ket{N
\uparrow}}\,,
\nonumber
\end{eqnarray}
where the electromagnetic current $J_{\alpha}^p$ is given by\cite{25} 

\begin{equation}
J_{\alpha}^p = \bar u_{N^*}(p') \left[ F_{1}^p(q^2) \left( q \!\!\!\!\, /
q_{\alpha} - q^2\gamma_{\alpha}\right) + F_{2}^p i \sigma_{\alpha \beta}
q^{\beta} \right] u(p)
\end{equation}
and $F_{1,2}^p(q^2)$ are the electromagnetic transition form factors for
the N-N$^*$ transitions, ${\bbox q}$ is
the three momentum of the virtual photon and $k_R$ is the energy of an
equivalent real photon both in the rest frame of N$^*$. They are given by

\begin{equation}
k_R=\frac{W^2-M^2}{2 W} \ , \ {{\bbox q}\,}^2 = \frac{\left( W^2-M^2+q^2
\right)^2}{4 W^2} - q^2 \ , \ W^2=(k + p)^2.
\end{equation}
 
\noindent
Using equations (21) and (22), isovector helicity amplitudes are derived to be

\begin{equation}
A_{\frac{1}{2}}^V(q^2) = {|\bbox q}|\, g(q^2) 
\left[ F_2^V(q^2) - \frac{q^2}{W+M}\, F_1^V(q^2) \right]\,,
\end{equation}

\begin{equation}
S_{\frac{1}{2}}^V(q^2) = \frac{1}{\sqrt{2}}\, {\bbox q}^2\, 
g(q^2) \left[ F_1^V(q^2) - \frac{F_2^V(q^2)}{W + M} \right]\,,
\end{equation}

\noindent
with

\begin{equation}
g(q^2) = \sqrt{\frac{8\pi \alpha (W+M) W^2}{M (W-M)((W+M)^2 - q^2)}}.
\end{equation}

\noindent
Inverting equations (25) and (26) we calculate the isovector form factors
$F_1^V(q^2)$ and $F_2^V(q^2)$ in terms of the helicity amplitudes and use
equation (21) to obtain them from the presently available data on
$A_{\frac{1}{2}}^p(q^2)$ and $S_{\frac{1}{2}}^p(q^2)$, quoted by Li {\it et
al.} \cite{28}.

In the case of axial vector form factors $F_A^V(q^2)$ and $F_P^V(q^2)$,
there is no experimental information available. We use the pion pole
dominance of the divergence of axial current (PDDAC) hypothesis, as done in
section II.B, to relate $F_A^V(q^2)$ and $F_P^V(q^2)$ to each other and also to
obtain a corresponding Goldberger Treiman relation (GTR) relating $F_A^V(0)$
to the $N^* \rightarrow N \pi$ coupling $\tilde{f}$ and $f_{\pi}$. 
A straightforward calculation gives \cite{24}

\begin{equation}
F_A^V(0) = \sqrt{2}\, f_{\pi}\, \frac{\tilde{f}}{m_{\pi}} \ , \ F_P^V = 
\frac{M + M^*}{m_{\pi}^2 -q^2}\, F_A^V\,,
\end{equation}

\noindent
where $\tilde{f}$ is the $N^* \rightarrow N \pi$ coupling determined
from the experimental decay rate  for this channel and defined through the
$N^*N\pi$ Lagrangian using pseudovector coupling, i.e. 

\begin{equation}
{\cal L}_{int} = i \frac{\tilde{f}}{m_{\pi}} \bar{\psi}_{N^*} \gamma^\mu 
\gamma_5 \bbox{\tau} (\partial_{\mu}\bbox{\phi})\psi  + \ h.c.  
\end{equation}

\noindent
A dipole form for the $q^2$ dependence of $F_A^V(q^2)$ is used

\begin{equation}
F^V_A(q^2)=\frac{F^V_A(0)}{(1 - q^2/M_A^2)^2}\,,
\end{equation}
where $M_A=1.0$ GeV as taken in the case of N-$\Delta$ form factors. 
Using Eqs. (28) and (30) for the axial vector form factor
$F_A^V(q^2)$ ( $F_P^V(q^2)$ contribution is negligible ) and Eqs. (25)-(27) 
for the vector form factor, numerical results are presented in the next 
section.

\section{Numerical Results}

In this section we present numerical results for the differential cross
section for the processes $e^- + p \rightarrow \Delta^0 + \nu_e$, $e^+ + p
\rightarrow \Delta^{++} + \bar{\nu}_e$, $e^- + p \rightarrow N^* + \nu_e$ and
study them using various form factors. We stress here, in particular, the
importance of the decay width in the angular dependence of the cross sections
and the effect of changing various form factors in the vector and axial vector
sectors.

\subsection{$e^- + p \rightarrow \Delta^0 + \nu_e$}

We present in Fig. 1 the $\Delta$ angular distribution for energies $E_e =
0.5,\, 0.855$ and $4.0$ GeV using the expressions for the form factors of N
-$\Delta$ transition, given in section II.B.  The electron
energies are chosen to correspond to the Mainz and TJNAF accelerators. 
The invariant mass has been restricted to $W < 1.4$ GeV to include the 
$\Delta$ dominated events only. The differential cross section is found to be 
forward peaked at lower electron energies, for example at $E_e=500$ MeV,
but the peak shifts to higher angles as we increase the
energy. There is a gain of $50\%$ in the total cross section 
as we go from the maximum Mainz energy (0.855 GeV) to 4 GeV 
We also study the cross section sensitivity to the transition
form factors. We do this by calculating the cross section for three sets of
vector and axial vector form factors. In the first set, we use Eqs. (12) and
(13) for the vector form factors and Eq. (18) for the axial vector form
factors, and the results are shown in Fig. 1 (solid line). In the second set, 
we take the form factors recently discussed by Hemmert {\it et al}.\cite{17},
which use Eqs. (12) and (14) for the vector
form factors and Eq. (18) with
$C_5^A(0)=0.87$, $C_4^A(0)=-0.29$ MeV for the axial vector form factors.
The results are shown by the short-dashed line. In the
third set, we use the nonrelativistic quark model form factors given by
equations (15) and (19) taken from Liu {\it et al.}\cite{16} and the results
are shown by the long dashed line. 

In Fig. 2, using the first set of form factors,
we show the effect of the decay width $\Gamma$ on the differential cross
sections $d\sigma / d\Omega_{\Delta}$ for $E_e=500$ and $4000$ MeV.
It is clear from Fig. 2 that the $\Delta$ width plays an important role in
the angular cross section. In the limit of $\Gamma \rightarrow 0$, our
results qualitatively agree with those of Hwang {\it et al.}\cite{15}.
The narrow angular range in
which the cross sections were earlier predicted to dominate is not there  
when the effect of decay width is taken into account. On the other hand there 
is a considerable cross section over a wide angular region, which increases as
energy raises and corresponds to $0 < \theta < 45^{\circ}$ for
$E_e = 4.0$ GeV. Therefore, a high angular resolution is not really needed 
in the experiments and large acceptance detectors can be
used to study this reaction. This feature of angular dependence of the cross
section is maintained with all the form factors used in this study.

\subsection{ $e^{+}+p \rightarrow \Delta ^{++}+ \bar\nu_{e} $      }

In Fig. 3, we present the results for $e^{+}+p \rightarrow \Delta ^{++}+
\bar\nu_{e}$. For this process, the cross section is overall enhanced 
by an isospin factor of 3 and  
reduced due to the different sign of the interference term, which depends on
energy and momentum transfer. The angular dependence of the cross section and 
its increase with the energy are, otherwise, quite similar to the $e^- + p
\rightarrow \Delta^0 + \nu_e$. 

The role of interference terms is very interesting in the case of N-
$\Delta$ transition. As a comparison of figures 1 and 3
shows, the suppression due to the opposite sign of interference terms is quite 
large at lower energies to overtake the overall increase by a factor of $3$ due 
to isospin. As the energy increases, the relative importance of the
interference terms becomes small and the cross section $e^{+}+p
\rightarrow \Delta ^{++}+ \bar\nu_{e}$ dominates. At around $E_{\nu} \sim
1.5$ GeV, the cross sections for $e^- + p \rightarrow \Delta^0 + \nu_e$ and 
$e^{+}+p \rightarrow \Delta ^{++}+ \bar\nu_{e}$ are comparable. The effect of 
the decay width of $\Delta$ is same as discussed in
section III.A, and our results with $\Gamma =  0$ are in qualitative 
agreement with the results of Hwang {\it et al.}\cite{15}, except that we 
obtain a larger cross section compared to the cross sections obtained by them
in the region away from the peak. 

\subsection {$e^{-}+p \rightarrow N^*+ \nu_{e}$}

In Fig. 4, we present the results for the $d\sigma / d\Omega_{N^*}$ at
$E_e = 4$ GeV with the same invariant mass cut for N$^*$ production as for
$\Delta$ production ( $W < 1.4$ GeV ). We use the form
factors obtained from Eqs. (25)-(27). In order to calculate them
we take the amplitudes $S_{\frac{1}{2}}^p$ and $A_{\frac{1}{2}}^p$ of
Gerhardt\cite{27} as quoted by Li {\it et al.}\cite{28}. The solid and
dashed curves shown in Fig. 4 correspond to the two parametrizations of
$S_{\frac{1}{2}}^p$ and $A_{\frac{1}{2}}^p$ of Ref.\cite{28}.
We find that the cross sections for N$^*$ production are smaller than the
$\Delta$ production cross sections by an order of magnitude. Furthermore, the
N$^*$ angular cross section peaks around $\cos \theta=0.82$ as compared to the
$\Delta$ production that peaks around $\cos\theta=0.73$.
The present uncertainty in the determination of the form
factors $F_1^V(q^2)$ and $F_2^V(q^2)$ leads to an uncertainty of $20 \%$
in the cross section in the peak region, as shown in Fig. 4. This uncertainty
does not affect the main conclusion of this study as the contribution of
N$^*$ production in the kinematic region of pions coming from $\Delta$ decay
is still too small and peaks at a different angle than the $\Delta$'s.
Increasing the invariant mass cut from $1.4$ GeV to $(M^* + m_{\pi})$, shifts
the peak of N$^*$ to still lower angles. Therefore, the larger width of the
N$^*$ affects pion production rates in an angular region well separated
from the $\Delta$ produced pions. 
\section{ Discussion and Conclusions}

We now address ourselves to the present experimental situation and the
possibility of observing these reactions at Mainz and/or TJNAF accelerators.
At these accelerators luminosities of the order of 
$10^{38} $cm$^{-2}$ sec$^{-1}$ or more are expected. The estimated count rate 
is given by

\begin{equation}
$counts$/$hour$ = \Delta\Omega \times \frac{d\sigma}{d\Omega} \times
$Luminosity$ \times 3600\, $sec$/$hour$ \times $Detector efficiency$.
\end{equation}
Using this formula, for example at 4.0 GeV, in the peak region of
around $40^\circ$ where the cross sections are of the order of $10^{-39}$
cm$^2$, we find the count rate to be

\begin{equation}
$counts$/$hour$ \sim 360 \times \Delta\Omega($sr$) \times $Detector efficiency$
/ $hour$.
\end{equation}
A similar count rate is expected at $E_e = 855$ MeV in the vicinity of
$20^\circ$, where the cross sections are of the same order. Keeping in mind
the finite angular range over which the cross sections are appreciably
larger than $10^{-40}$ cm$^2$, the estimates made above suggest that the number
of counts could be high enough for considering the feasibility of doing such
an experiment.

Finally, to summarize the paper, we have made a theoretical study of
the weak production of $\Delta$ and N$^*(1440)$
through the charge changing reactions induced by electron beams of
energies corresponding to Mainz and TJNAF accelerators. We find that :

\begin{enumerate}
\item The differential cross section for the weak production of $\Delta$
resonance with electron beams of the order of $10^{-39}$
cm$^2$/sr which is quite sizeable. At $E_e=855$~MeV, the cross 
section for $e^- + p \rightarrow \Delta^0 + \nu_e$ is larger than the cross 
section for $e^+ + p \rightarrow \Delta^{++} + \bar{\nu}_e$ while at 
$E_e=4.0$~GeV, the cross section for $e^+ + p \rightarrow \Delta^{++} + 
\bar{\nu}_e$ process is about a factor two larger than the 
$e^- + p \rightarrow \Delta^0 + \nu_e$. As we increase the energy from $855$ 
to $4000$ MeV the peak in the cross section shifts to a higher angle from 
$20^{\circ}$ to $40^{\circ}$.
\item  There is a large angular region in which the differential cross
sections are appreciable. This feature of the differential cross sections
facilitates the observation of this reaction at current electron
accelerators where large angle acceptance detectors are planned to be used
in electron scattering experiments. There is no need for a
sharp angular resolution in the vicinity of $0.1^{\circ}$ as found
earlier, based on a calculation neglecting the decay width of $\Delta$
resonance.
\item The production cross section for N$^*$ is an order of magnitude
smaller than the production cross section of $\Delta$ and peaks at an angle
well separated from the $\Delta$ production peak region. This makes the
identification of $\Delta^0$ through the measurement of pions and protons
quite clean for the invariant mass cut of $W < 1.4$ GeV. There is
no such contamination from N$^*$ resonances in the identification of
$\Delta^{++}$.

\item The production cross section is dominated by the three form factors
$C_5^A$, $C_3^V$ and $C_4^V$ and an experimental measurement could
discriminate between the various models used for these form factors. If the
electromagnetic production cross sections for the $\Delta$ resonance are
precise enough to fix the $C_3^V$ and $C_4^V$ form factors, this will
make the determination of $C_5^A$ quite model independent.

\item There is a very strong energy dependence of the V-A interference terms
in $\Delta$ production process which can be used for determining the other
form factors like $C_3^A(q^2)$ and $C_4^A(q^2)$. An experimental information
on these form factors is extremely important for the theoretical models
currently used for nucleon structure as well as for some earlier analyses
which use quite different values of $C_3^A$ and $C_4^A$ for explaining the
experimental data on neutrino scattering in the intermediate energy region.
\end{enumerate}

\acknowledgments

We wish to acknowledge useful discussions with F. Cano, G. A. Miller and A.
Joshipura. This work has been partially supported by DGYCIT contract no.
AEN-96-1719. One of us (L.A.R.) acknowledges financial support from the
Generalitat Valenciana. S.K.S. has the pleasure of thanking prof. Oset for
his hospitality at the University of Valencia and acknowledges financial
support from the Ministerio de Educaci\'on y Cultura of Spain in his
sabbatical stay.

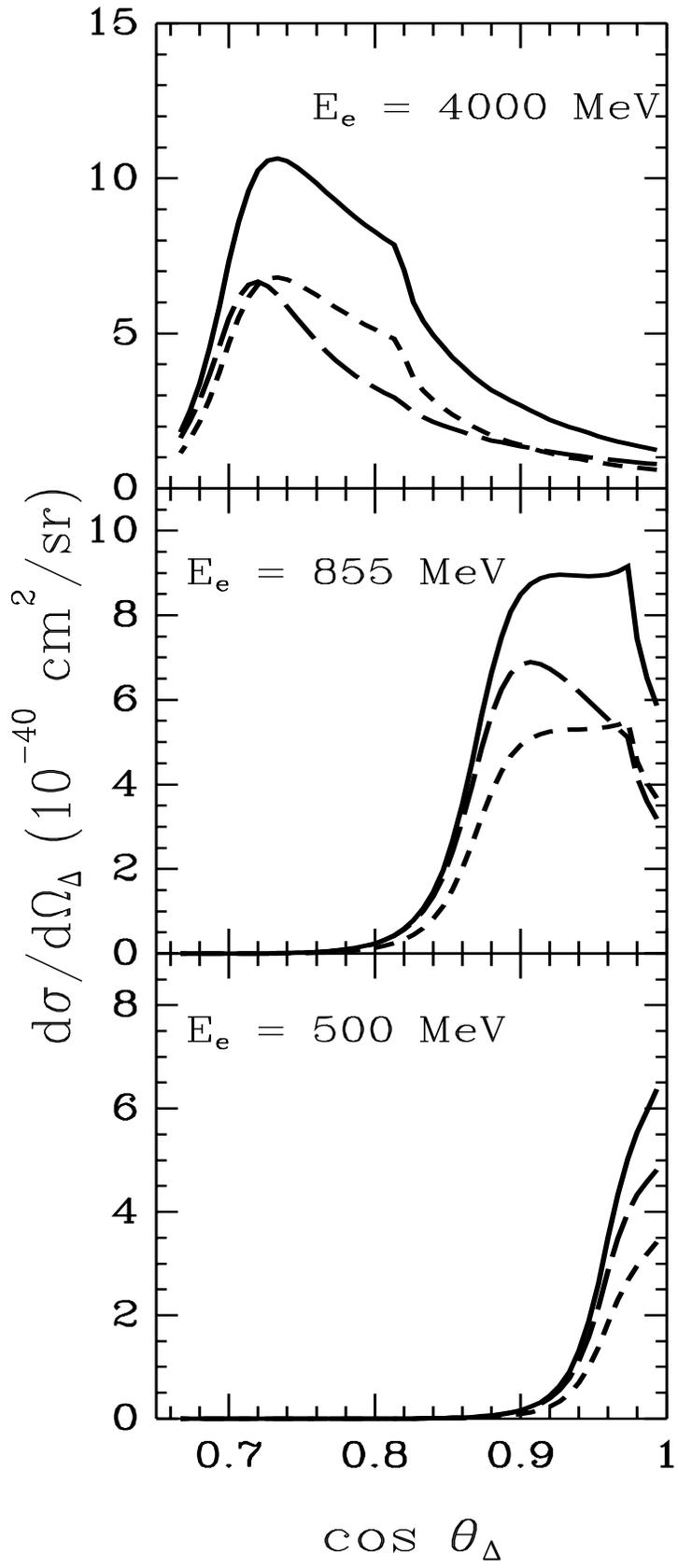
\begin{figure}
\caption{$\Delta^0$ angular distribution for the reaction $e^- + p
\rightarrow \Delta^0 + \nu_e$ with three different sets of form 
factors, as explained in the text.}
\end{figure}

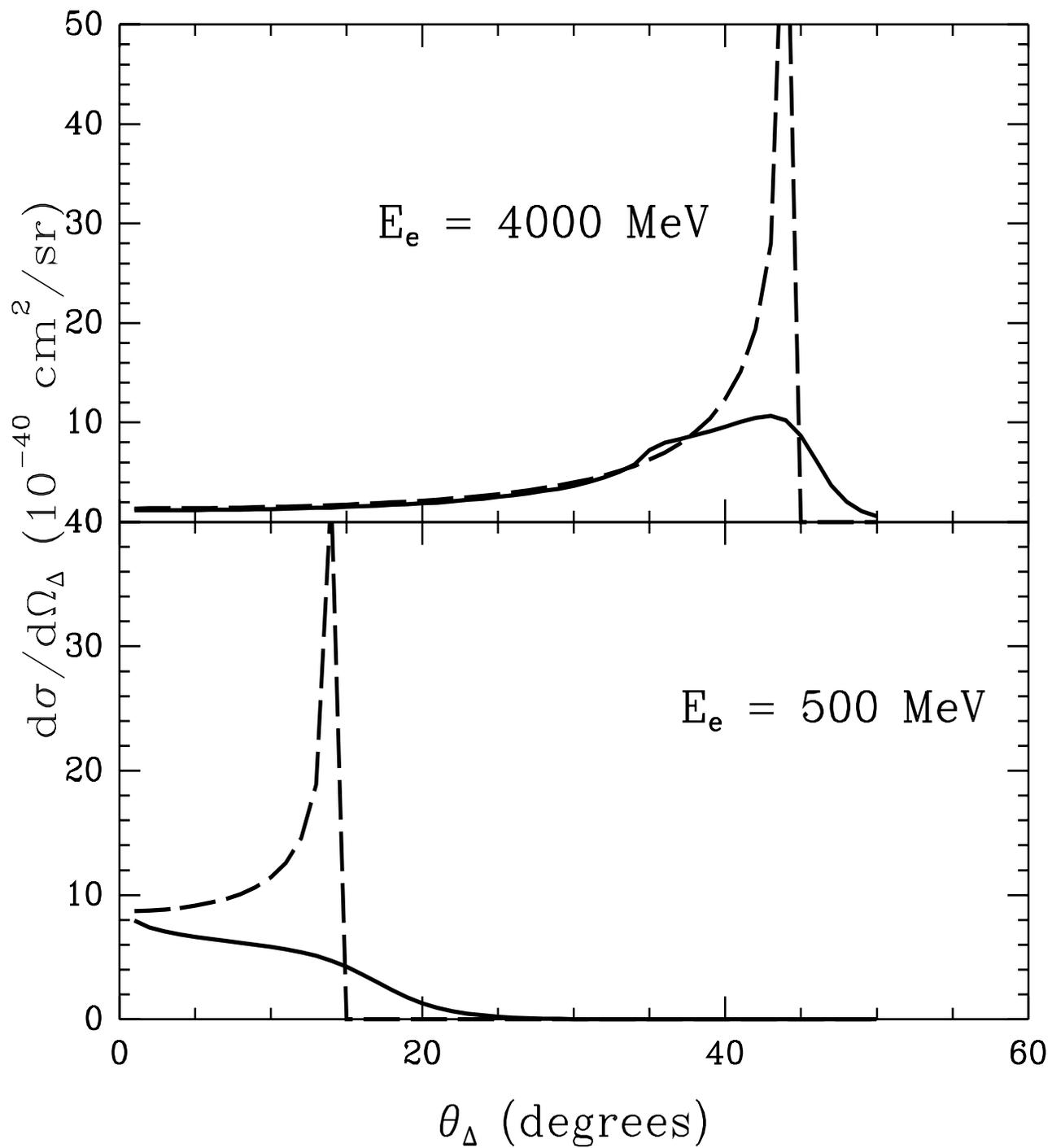
\begin{figure}
\caption{$\Delta^0$ angular distribution for the reaction $e^- + p
\rightarrow \Delta^0 + \nu_e$ with finite (solid line) and zero (dashed line) 
widths. The form factors are taken to be the same as for the solid line of
Fig. 1.}
\end{figure}

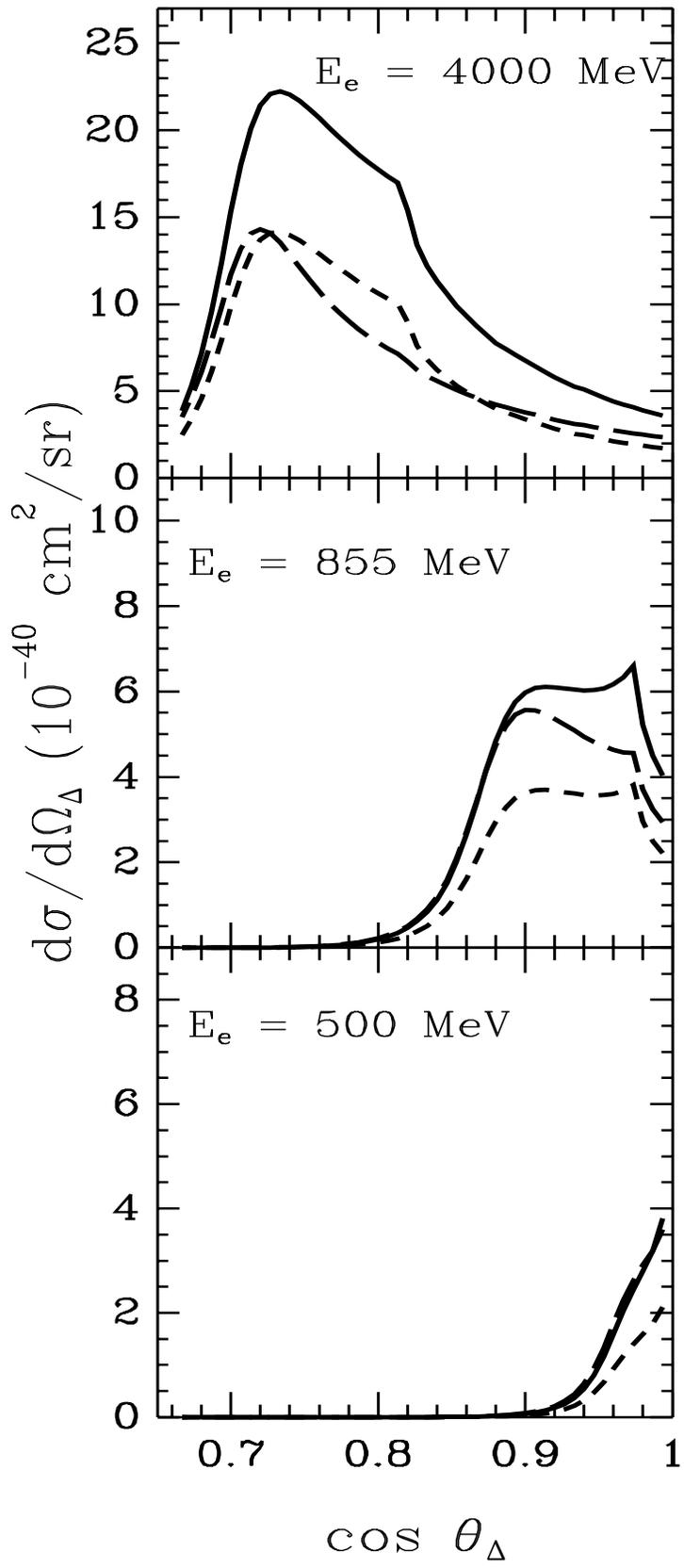
\begin{figure}
\caption{Same as Fig. 1 for the reaction 
$e^+ + p \rightarrow \Delta^{++} + \bar\nu_e$.}
\end{figure}

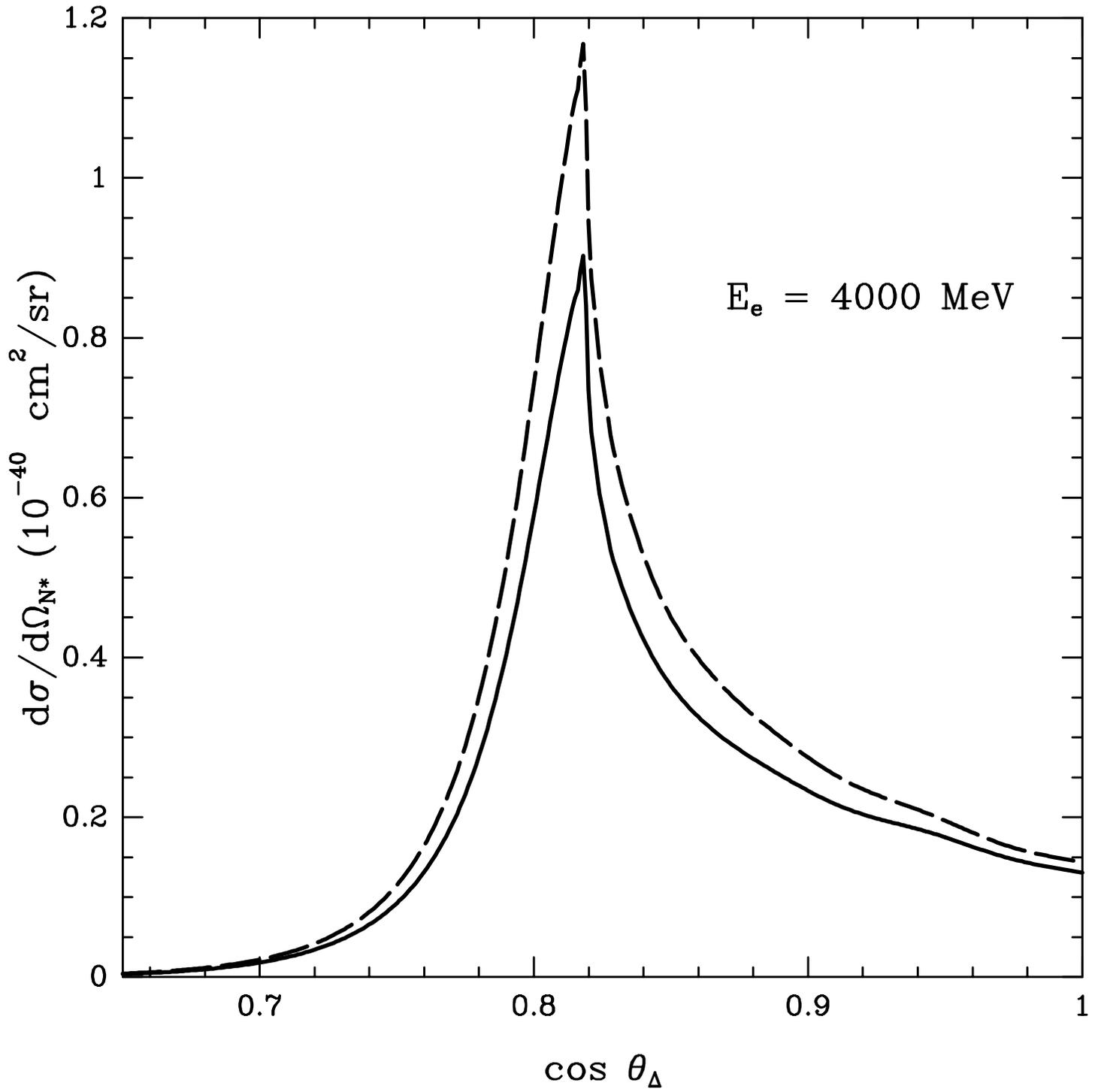
\begin{figure}
\caption{N$^*$ angular distribution for the reaction $e^- + p \rightarrow
N^* + \nu_e$ with two different parametrizations of the vector form factors 
extracted from Gerhardt's analysis [27], as explained in the text.}
\end{figure}

\end{document}